\documentstyle[11pt,epsf]{article}
\begin{document}
\thispagestyle{empty}
\baselineskip 15pt
\rightline{KIAS-P02022}
\rightline{SNUTP-02009}
\rightline{{\tt hep-th/0204109}}
\

\def\tr{{\rm tr}\,}
\def\Tr{{\rm Tr}\,}
\newcommand{\beq}{\begin{equation}}
\newcommand{\eeq}{\end{equation}} \newcommand{\beqn}{\begin{eqnarray}}
\newcommand{\eeqn}{\end{eqnarray}} \newcommand{\bde}{{\bf e}}
\newcommand{\balpha}{{\mbox{\boldmath $\alpha$}}}
\newcommand{\bsalpha}{{\mbox{\boldmath $\scriptstyle\alpha$}}}
\newcommand{\betabf}{{\mbox{\boldmath $\beta$}}}
\newcommand{\bgamma}{{\mbox{\boldmath $\gamma$}}}
\newcommand{\bbeta}{{\mbox{\boldmath $\scriptstyle\beta$}}}
\newcommand{\btau}{{\mbox{\boldmath $\tau$}}}
\newcommand{\lambdabf}{{\mbox{\boldmath $\lambda$}}}
\newcommand{\bphi}{{\mbox{\boldmath $\phi$}}}
\newcommand{\bslambda}{{\mbox{\boldmath $\scriptstyle\lambda$}}}
\newcommand{\ggg}{{\boldmath \gamma}} \newcommand{\ddd}{{\boldmath
\delta}} \newcommand{\mmm}{{\boldmath \mu}}
\newcommand{\nnn}{{\boldmath \nu}}
\newcommand{\diag}{{\rm diag}}
\newcommand{\bra}{\langle}
\newcommand{\ket}{\rangle}
\newcommand{\sn}{{\rm sn}}
\newcommand{\cn}{{\rm cn}}
\newcommand{\dn}{{\rm dn}}
\newcommand{\tA}{{\tilde{A}}}
\newcommand{\tphi}{{\tilde\phi}}
\newcommand{\bpartial}{{\bar\partial}}
\newcommand{\br}{{{\bf r}}}
\newcommand{\bx}{{{\bf x}}}
\newcommand{\bk}{{{\bf k}}}
\newcommand{\bq}{{{\bf q}}}
\newcommand{\bQ}{{{\bf Q}}}
\newcommand{\bp}{{{\bf p}}}
\newcommand{\bP}{{{\bf P}}}
\newcommand{\thet}{{{\theta}}}
\newcommand{\tauu}{{{\tau}}}
\renewcommand{\thefootnote}{\fnsymbol{footnote}}
\newcommand{\E}{{{\cal E}}}
\newcommand{\N}{{\scriptscriptstyle N}}

\vskip 0cm
\centerline{ \Large \bf Tales of D0 on D6 Branes:}
\centerline{\Large\bf  Matrix Mechanics of Identical Particles} 
\vskip .2cm

\vskip 1.2cm
\centerline{Chanju Kim$^a$\footnote{Electronic Mail: cjkim@phya.snu.ac.kr},
Kimyeong Lee $^{b}$\footnote{Electronic Mail:
klee@kias.re.kr}, and Sang-Heon Yi$^{a}$\footnote{Electronic Mail:
shyi@phya.snu.ac.kr}
}
\vskip 10mm
\centerline{ \it $^a$ Department of Physics and Center for Theoretical Physics}
\centerline{ \it Seoul National University, Seoul 151-742, Korea}

\vskip 3mm
\centerline{ \it $^b$ School of Physics, Korea Institute for Advanced
Study}
\centerline{ \it 207-43, Cheongryangri-Dong, Dongdaemun-Gu, Seoul
130-012, Korea
}
\vskip 1.2cm


\begin{quote}
{We investigate  a class of matrix model which describes the dynamics of
identical particles in even dimentional space.  We show that the 
degrees of freedom after some constraints are implimented is proportional
to particle number and consist of those for positions and internal
degrees.  The particle dynamics is given by the metric on the smooth
moduli space. The moduli space metric for two particles is  found. The
size of tightly packed $N$ particles grows like $\sqrt{N}$.  Our
matrix model is related to the matrix model for fractional quantum
Hall effect, the ADHM formalism of $U(1)$ instantons on noncommutative
space, and supersymmetric D0 branes on D6 branes with nonzero B-field
in type IIA theory.}
\end{quote}


\newpage
\baselineskip 15pt


Recently, there have been some studies to understand the dynamics of
identical particles in terms of the matrix gauge theory.  For $N$
particles in $D+1$ dimensional space, the particle coordinates become
$D+1$ square matrices of size $N$. The particle exchange symmetry
becomes a part of the gauge symmetry and so is included in the
dynamics classically. Usually the matrix theory has more degrees of
freedom than ones necessary for particle dynamics and further
constraints are imposed to reduce the degrees of freedom.  A typical
example is the dynamics of instanton in terms of the matrices in the
ADHM construction~\cite{adhm}. These matrices are constrained by the
three D-term constraints and by the Gauss law and so the number of net
degrees of freedom is proportional to the instanton
number~\cite{adhm,wittenadhm}.  The number of degrees of
freedom for a single instanton is the sum of that for the position and
that for the internal degrees. Another example is the Chern-Simons
inspired matrix model which describes the fractional quantum Hall
fluid.~\cite{Susskind,poly}.

In  this work  we  study  a matrix  theory  which describes  identical
particles in even dimensional space. Our matrix theory is quadratic in
time  derivative  and   so  naturally  describes  the  nonrelativistic
Newtonian  dynamics of  identical particles.  We count  the  number of
degrees of freedom left after the constraints are imposed, which turns
out to  be proportional  to particle number.  For the  relative moduli
space for two  particle case, we extend our  result on four dimensions
to other  even dimensions.  The  size of tightly packed  $N$ particles
grows like $\sqrt{N}$.

Our theories appear in many context.  On two dimensional space it is
somewhat related to the first order matrix theory introduced by
Susskind and Polychronakos. On four dimensional space with single
flavor it is related to the ADHM formalism of $U(1)$ instantons on
noncommutative four space.  On six dimensional space it describes the
low energy dynamics of supersymmetric D0 branes on D6 branes, which is
possible with nontrivial background $NS$ $B$ field in type IIA theory
as discussed by Witten~\cite{d0d6}. Some of issues studied here has
been discussed in somewhat simplier level in Ref.~\cite{ohta}.

The ADHM description of the low energy dynamics of instanton is a
gauge theory of matrices with additional constraints, which can be
solved in terms of the moduli parameter. Once the constraints are
solved and the gauge degrees of freedom are  removed, the instanton
moduli space metric describes the motion of instantons. The metric on
the moduli space is singular and the instanton dynamics is incomplete.

The ADHM formalism arises naturally from D0 brane on D4 brane point of
view. Its low energy dynamics of $D0$ brane is given by the Yang-Mills
matrix mechanics model with eight supersymmetries. The relevent fields
are adjoint and fundamental scalar fields.  These matrices are
constrained in the ADHM formalism by the three D-term
potentials~\cite{adhm,wittenadhm}.

When nonzero background  $B$ field introduced on D4 branes, their
field theory becomes Yang-Mills theory on noncommutative
space~~\cite{CDS,Seiberg}, which can be obtained from a certain limit
of open string theory on D4 branes with background B-field. In this
case, one can have nonsingular moduli space for
instantons~\cite{ns}. (See Ref.~\cite{dn,nek} for a recent review.)
The B-field appears as the FI term in the matrix theory, which blows up
the singularities of the instanton moduli space with a finite scale.
The ADHM formalism is  simplified  when the gauge group is
$U(1)$~\cite{Nakaj,Furuuchi}.  Our class of the matrix model includes
only the ADHM formalism for the $U(1)$ group.

In two dimensional space Susskind and Polychronakos proposed and
studied the matrix model for charged particles on the lowest Landau
level and so their Lagrangian is first order in time derivative and
found to be related to the noncommutative Chern-Simons
theory~\cite{Susskind,poly}. The space of matrices become a classical
phase space. After the Gauss law is imposed and the gauge degrees are
removed, the remaining degrees of freedom are  proportional to the
number of particles. Especially the quantum mechanical eigenvalues and
eigenfunctions of a quadratic  Hamiltonian can be found
exactly~\cite{heller}.

On the other hand our matrix model is second order in
time derivative and so it describes the real time evolution of
particles.  After reducing the degrees of freedom, the motion of
identical particles is  described by the coordinates of the 
moduli space. The moduli space, which is  the coordinates space
in our model, appears as the phase space in the matrix model of
Susskind and Polychronakos.

In the type IIA theory, D0 branes feel repulsive force from the
parallel D6 branes. However when there exists a uniform background B
field above critical strength, D0 branes can be attracted to D6 branes
and form a BPS configuration. The  D0 brane physics is
described by the matrix model~\cite{d0d6}.
This is exactly the matrix model we study in the six dimensions.

Noninteracting identical particles on flat $R^D$ have a singular
configuration space $(R^D)^N/S_N$ where $S_N$ is the permutation group
of $N$ particles. The configuration space is singular when
particles come together. On even dimensional space our matrix model is
a natural blow-up procedure of these singularities.  There is
a natural length scale for this blow up, making particles to carry
effectively finite size core.  When many identical particles described
by the matrix come together, their kinetic energy, which determines
the moduli space metric, makes them to behave somewhat like incompressible
fluid even though there is no force between them.  In our case the
total volume occupied by  tightly packed  $N$ particles in $D=2k$ dimensions
grows as $N^k$, which is faster than the incompressible gas except
$k=1$. As we will see, the kinetic energy also entails the long range
interaction between particles. Thus, one could regard the matrix model
to describe a peculiar class of particles living on even dimensional
space.

We start with the matrix model description applicable for all these
cases. The matrix model for $N$ particles on $2k$ dimensional space
is described by $N\times N$ complex matrices $Z_i$ with $i=1,...,k$ and
$N\times M$ complex matrix $\psi$. The kinetic term for these matrices
is 
\beq
{\cal L} = \tr \left( D_0Z_i D_0\bar{Z}_i + D_0\psi D_0\bar{\psi}
\right) .
\label{lagr}
\eeq
The $U(N)$ local gauge transformation leads to $Z_i\rightarrow U Z_i
\bar{U}$, $\psi\rightarrow U\psi $ and $A_0 \rightarrow U A_0 \bar{U}
- i \partial_0 U \bar{U} $. The $U(M)$ global flavor symmetry leads to
$\psi \rightarrow \psi \bar{V} $. The matrices $Z_i,\psi$ have $2kN^2
+2NM$ real parameters. Out of which $N^2$ are gauge parameters,
leading to $(2k-1) N^2+2NM$ physical parameters, which is clearly much
larger than the number of particles. The spatial rotational group
$SO(2k)$ is broken to $U(k)$.  (See for a recent investigation of the
model with $k=1/2$ and so particles living on a line where the matrix
degrees of freedom are proportional to the particle
number~\cite{jhp}.)

Additional constraints are needed for these matrices. We choose them to be
\beqn
&&\sum_{i=1}^k [Z_i,\bar{Z}_i]+\psi\bar{\psi}=\zeta 1_N, 
\label{kq} \\
 &&[Z_i,Z_j]=0 \quad i,j=1,2,...,k    \;.
\label{fterm},
\eeqn
These can be regarded as the minimum of a potential of matrices.  In
the theory with four supersymmetries the first one (\ref{kq}) can be
regarded as the minimum condition of the $D$ term potential and the
second one as the minimum condition of the $F$ term potential.

For the two dimensional case, there is only one constraint (\ref{kq}).
The number of the maximal supersymmetry is four and the constraint is
the minimization condition of the D-term potential.  For the four
dimensional case with a single flavor $M=1$, the above constrains are
reduced version of the ADHM constraints for the $U(1)$ instantons on
noncommutative four space, and so the number of underlying
supersymmetry is eight.  For the theories of the six dimensional
system, the number of maximal supersymmetry is again four and the
constraint $(\ref{fterm})$ comes from the superpotential $\tr
Z_1[Z_2,Z_3]$. For other cases there is no obvious supersymmetry is 
as there is no suitable superpotential for the constraint
(\ref{fterm}).

For one particle with $N=1$ Eqs~(\ref{kq}) and (\ref{fterm}) have the
trivial solution where $Z_i$ are arbitrary numbers and
\beq 
\psi\bar{\psi}=\zeta1_N .
\label{psieq}
\eeq
The $\psi$ space is $S^{2M-1}$ and becomes $CP^{M-1}$ after we mod out
by gauge symmetry $U(1)$.  The particle position $Z_i$ has $2k$ parameter
and the  internal degrees of freedom has $2(M-1)$ parameters. The
total degrees of freedom for a single particle is then $2k + 2M -2$.

The supersymmetric quantum dynamics of particles with
$M\ge 2$ is possible on six dimensions. The quantum mechanics of the
sigma model with moduli space metric has $4$ 
supersymmetry as the maximal supersymmetry. For a single particle with
nontrivial internal degrees of freedom, there are $M$ quantum
mechanical ground states because they are normalizable harmonic
forms~\cite{Witten} and their total number can be identified with the
Euler number of $CP^{M-1}$ by Hodge theorem~\cite{egh}.

When we consider many particles $N\ge 2$, one may wonder how degrees
of freedom are there. Naively we expect that each particle carries the
same number and so the total number of free parameters is the sum of
that for constituent particles.  However it is not obvious at all from
Eqs.~(\ref{kq}) and (\ref{fterm}) imposed on the matrices $Z_i, \psi$.
Eq.(\ref{fterm}) seems overconstraining. Indeed the constraints in
Eq.~(\ref{fterm}) are not independent. In Appendix A we show that the
correct counting leads to the $N(2k+2M-2)$ degrees of freedom as
expected.  The original kinetic term then defines a K\"ahler space of
dimension $N(2k+2M-2)$ with the induced K\"ahler form from the flat
K\"ahler form corresponding to the kinetic energy. Our matrix model is
a natural method to define a class of K\"ahler spaces, which is related
to the blow-up of the configuration space of identical particles.

It is very hard to solve the constraints for arbitrary number of
particles $N$ and the internal degrees of freedom characterized by
$M-1$.   When $M\ge 2$, there are some internal degrees
of freedom and so one can overlap the particles at a same
point. Especially with $M\ge N$, one can solve Eqs.~(\ref{kq}) and
(\ref{fterm}) with $Z_i=0$ and the $\psi$ space satisfying
Eq.~(\ref{psieq}) becomes Grassmannian manifold $U(M)/U(N)\times
U(M-N)$ after moding out the gauge group, as identified by
Witten~\cite{d0d6}. When there is no internal degrees of freedom,
$N$ particles act like having a  `hard core', given not by force 
but by the metric of the kinetic term. When there are enough internal
space $M\ge 2$,  particles of number less than or equal to $M$ can
come together to the top on each other. 

On two dimensional space $k=1$ with $M=1$, the most general solution
of the constraint (\ref{kq}) has been known before. (See for example
Ref.~\cite{poly}.)  A further discussion of our model on two
dimensions and its relation to Polychronakos system is given in
Appendix B.  In four dimensional space the solution for two particle
case has been found before~\cite{tong} and has been used by us to show
the moduli space metric to be the Eguchi-Hanson metric. In the
following we extend this analysis to arbitrary even dimensions.

With no internal degrees of freedom $M=1$ and so there is only single 
flavor $\psi $. In this case one can find the generic two particle
solutions ($N=1$) for these constraints by triangularizing the commuting
complex matrices $Z_i$ simultaneously by unitary transformations,
\beqn 
Z_i=w_i 1_2+ {z_i\over 2}\left ( \begin{array}{cc} 
1 & \sqrt{\frac{2b}{a}} \\ 
0 & -1 \end{array} \right ),\ \psi=\sqrt{\zeta}\left(\begin{array}{c}
\sqrt{1-b}
 \\ \sqrt{1+b} 
\end{array}\right),
\label{solution}
\eeqn 
where 
\beq 
a=\frac{\sum_i |z_i|^2}{2\zeta} \ge 0 , \;\;\; 
 b = \frac{1}{a +\sqrt{1+a^2}}  .
\eeq
Notice that the remnant $U(1)^2$ gauge is used for fixing the phase of
$a$ and $b$.  We can identify $w_i$ and $z_i$ as the gauge invariant
position of two particles at all separation of two particles not only
at large separation since $w_i\pm z_i/2$ are the eigenvalues of $Z_i$.
There is still the remaining discrete gauge symmetry
\beq
U=\left ( \begin{array}{cc} -b & \sqrt{2ab} \\ \sqrt{2ab} & b
\end{array} \right), 
\eeq
which changes the sign of $z_i$, leaving $\psi$ invariant.  This
discrete gauge symmetry identifies the positions of two identical
particles and making the relative moduli space at large separation to
be $R^{2k}/Z_2$. The orbifold singularity at the origin gets blow-up
in the matrix theory. At the coincident limit $z_i=0$, the space of
the solution (\ref{solution}) becomes a submanifold $S^{2k-1}$. We
have to mode out this by the U(1) gauge transformation ${\rm
diag}(e^{i\beta}, 1)$ because this transformation gets restored at
this limit $\sqrt{1-b}=0$.  Thus, the submanifold of moduli space at
the coincident limit is $S^{2k-1}/U(1)$ which can be identified with
$CP^{k-1}$ because $S^{2k-1}$ is $U(1)$ fiber bundle over $CP^{k-1}$
as we know~\cite{bais}.  Here $CP^0$ denotes just one point.

To obtain the tangent vectors on the moduli space, we consider the
infinitesimal variation of the above solution made of the
infinitesimal change in moduli parameters and the infinitesimal gauge
transformation,
\beq
\delta Z_i=d Z_i-i[\delta \alpha,Z_i],\
\delta \psi=d\psi-i\delta\alpha \psi \; .
\label{variation}
\eeq
We need to fix the gauge variation of the above solutions under small
variation, which can be achieved by solving the background gauge
condition. The initial configuration for the matrices is the solution
(\ref{solution}) and its `initial velocity', which is characterized by
the variation. It should satisfy the Gauss law constraint, which is
identical to the background gauge fixing condition, 
\beq \sum_{i=1}^k([\delta Z_i,\bar{Z}_i]-[Z_i,\delta
\bar{Z}_i])+\delta \psi\bar{\psi}-\psi\delta\bar{\psi}=0 \; .
 \eeq
This condition fixes the infinitesimal gauge transformation  uniquely,
\beqn
&&\delta \alpha= \frac{ib}{2\sqrt{1+a^2}} \left( \begin{array}{cc}
\frac{\bar{\partial} a -\partial a}{1-b} & -\frac{\sqrt{2} \bar{\partial} a}{\sqrt{ab}} \\
\frac{\sqrt{2}\partial a}{\sqrt{ab}} &  \frac{\bar{\partial}
a-\partial a}{1+b}  
\end{array} 
\right) ,
\eeqn
where $\partial a = \bar{z}_i dz_i/(2\zeta)$ and $\bar{\partial} a =
z_i d\bar{z}_i/(2\zeta)$.  Knowing the tangent vector $(\delta
Z_i,\delta \psi )$, we can obtain the metric in the usual way
\beq
ds^2=  \tr  \left( \sum_{i=1}^k\delta Z_i\delta
\bar{Z}_i+\delta\psi\delta\bar{\psi} \right) .
\eeq
Incidentally, $\delta \psi=0$ in our two particle case.
Then, the moduli space metric becomes 
\beq
ds^2=ds^2_{cm}+ds^2_{rel}, 
\eeq
where the center of mass and relative metrices are
\beqn
&&ds^2_{cm}= 2\sum_{i=1}^k dw_id\bar{w}_i,  \\
&&ds^2_{rel}=\frac{\sqrt{r^4+ 4\zeta^2}}{2r^2}\sum_{i=1}^k dz_id\bar{z}_i
 - \frac{2\zeta^2}{r^4\sqrt{r^4+4\zeta^2}} 
\sum_{i,j=1}^k\bar{z}_i z_j dz_i d\bar{z}_j  
\label{rel2}
\eeqn
with $r^2 = \sum_i |z_i|^2$.  At the large $r=\sqrt{\sum_i |z_i|^2} $
limit the metric approaches that of the flat space with the correction
of order $1/r^2$, which is a sign of long-range interaction.  The
above metric $ds^2_{rel}$ for any $k$ is a K\"ahler metric as 
$g_{i\bar{j}}=\partial_i\bpartial_j{\cal K}$ with the K\"ahler
potential, 
\beq
{\cal K}_{rel}={1\over 2}\sqrt{r^4+4\zeta^2} + \frac{\zeta}{2} \ln 
\frac{ \sqrt{r^4+4\zeta^2}-2\zeta}{\sqrt{r^4+4\zeta^2} +2\zeta}  .
\eeq
The K\"ahler form could be obtained by $K = {i\over
2}\partial\bpartial {\cal K}$.  This metric has $U(k)$ holomorphic
isometry transforming $z_i\rightarrow U_{ij}z_j$, which is a subgroup
of the spatial rotation $SO(2k)$.  In fact, the metric of the $k=2$
case has hyperK\"ahler structure and is  identified with the standard
Eguchi-Hanson manifold~\cite{tong}, where the isometry $SU(2)\subset
U(2)$ becomes triholomorphic. Basically the $k=1,3$ cases are related
to supersymmetric theory of four real supercharges but the $k=2$ case
is related to theory of eight real supercharges.  Therefore, the
metric of the $k=2$ case has the additional structure and is Ricci
flat.  Note that the metrics of the $k=1,3$ cases are not non Ricci
flat K\"ahler.

To explore the region near the origin of the relative metric $r \simeq
0$ where two particles come together, we first note that the
Fubini-Study metric on $CP^{k-1}$ can be given by
\beq
  ds^2_{FS}={1\over r^2} \biggl( \sum_idz_id\bar{z}_i -
\frac{1}{r^2}\sum_{i,j}\bar{z}_i z_j dz_i d\bar{z}_j \biggr) .
\label{FS}
\eeq
After  the following transformation $z_i=\lambda q_i$, $
i=1,2,...,k-1, $ and $z_k=\lambda $ with $ \rho\equiv
\big(\sum_{i=1}^{k-1}|q_i|^2 \big)^{1/2}$, 
the above metric (\ref{FS}) becomes the standard  one  on $CP^{k-1}$, 
\beq
 ds^2_{FS}= \frac{\sum_{i=1}^{k-1}dq_id\bar{q}_i}{1+\rho^2}
                  -\frac{\sum_{i,j=1}^{k-1}
                     \bar{q}_iq_jdq_id\bar{q}_j}{(1+\rho^2)^2}.
\eeq
The  metric $ds_{FS}^2$ is normalized so that the Ricci tensor
satisfies $R_{i\bar{j}} = 2k \delta_{i\bar{j}}$. The $CP^{k-1}$ space
is K\"ahler with K\"ahler form ${\cal K} $ which is locally exact,
${\cal K} = d{\cal A}$. On $R^{2k}$, $\sum_i dz_id\bar{z}_i = dr^2 +
r^2 d\Omega_{2k-1}^2$ with $\Omega_{2k-1}^2$ being the metric on the
unit $S^{2k-1}$.  The unit $S^{2k-1}$ sphere can be identified with
$U(1)$ fiber bundle over $CP^{k-1}$ and its metric
becomes~\cite{bais} 
\beq
d\Omega_{2k-1}^2 = (d\theta + {\cal A})^2 + ds_{FS}^2.
\eeq
Thus, the relative metric (\ref{rel2})  becomes
\beq
ds_{rel}^2= \frac{r^2}{2\sqrt{r^4+4\zeta^2 }} \biggl(dr^2+ r^2 
(d\theta+ {\cal A})^2 \biggr)
+{1\over 2} \sqrt{r^4+4\zeta^2} \; ds_{FS}^2 ,
\eeq
where the range of
$\theta$ is $[0,\pi]$ instead of $[0,2\pi]$ as   we have identified  $z_i$
with $-z_i$.  
With change of variable  $v=r^2$ the above metric becomes $ds^2 \approx
(dv^2+v^2 d(2\theta)^2 )/(8\zeta) + 2\zeta ds_{FS}^2$ near $r\approx
0$.  Thus, the above metric becomes the smooth metric of
$R^2\times CP^{k-1}$ near the origin. 

By the coordinate transformation 
\beq
u^4={r^4\over 4}+\zeta^2 ,
\eeq
we can put the metric and K\"ahler potential in the following form
\beqn
ds^2_{rel}&=& \frac{du^2}{1-{\zeta^2\over u^4}}
            +u^2\Big( ds^2_{FS}+(1-{\zeta^2\over u^4})
              (d\theta+{\cal A})^2\Big) ,    \\
{\cal K}_{rel}&=&u^2+{\zeta\over 2} \ln
(\frac{u^2-\zeta}{u^2+\zeta}) . 
\eeqn

 For the $k=2$ case there exists unique self-dual middle harmonic form
which corresponds to the supersymmetric normalizable ground state of
the maximal eight supersymmetric extension~\cite{tong} provides the
wave functions of the threshold bound state of two $U(1)$ instantons
on noncommutative four space.  For the $k=1,3$ cases, it is not clear
whether there exists any normalizable harmonic forms which correspond
the supersymmetric normalizable ground states of the maximal four
supersymmetric extension. 

Having studied two particle case, let us consider briefly $N$ particle
case with $M=1$. As all $Z_i$ are commuting due to the constraint
(\ref{fterm}), we can put all $Z_i$ as upper triangular matrices
simultaneously. {}From the analogue with two particle system, we expect
that all diagonal elements of $Z_i$ vanishes when all particles are
packed tightly together. (Here we put the center of the mass position
at the origin.)  Then we can find the solution of two constraints
(\ref{kq}) and (\ref{fterm}). Only nonvanishing components of $Z_i$ are
\beqn
&& (Z_i)_{a,a+1} = \sqrt{a\zeta} u_{ai} \;\;\; a=1,2,...,N-1, \\
&&  \psi_N=\sqrt{N\zeta},
\eeqn
where $u_{a}$ is $k$-dimensional unit complex vector.  We have used
one of the diagonal $U(1)$'s subgroups to make $\psi$ vector
real. Further use of the diagonal $U(1)$ subgroups leads to
identification $u_a \sim e^{i(\theta_a-\theta_{a+1})}u_a$, making each
$u_a$ vectors to belong to $CP^{k-1}$. Thus the solution  space of $N$
particles for the relative motion becomes $(CP^{k-1})^{N-1}$ when all
particles come together. For this solution
\beq
\tr Z_i \bar{Z}_i = \sum_{a=1}^{N-1} a \zeta = \frac{N(N-1)}{2} \zeta .
\eeq
While the particles are indistingushable, we could regard the radius
square of the a-th particle is $(a-1)\zeta$.  The radius of the
outmost particle is $\sqrt{(N-1)\zeta}$ and so the total volume in
$2k$ dimension will be proportional to $N^{k}\zeta^k$ in large $N$
limit.  For particles in two dimensions, the minimum volume grows like
$N$ which shows that they are incompressible. For particles in higher
dimensions $(D=2k)$, the volume grows faster than $N$, showing the
existence of additional mechanism.  It shows nontrivial interaction
between particles in short distance. Such  phenomena is not new: the
volme of nonabelian core of tightly packed $N$ BPS monopoles grows as
$N^3$ in three dimensions.  The solution space of the packed
particles, $(CP^{k-1})^{N-1}$ is trivial for the two dimensional case
and so the configuration is rotationally invariant. That is nontrivial
for the higher dimensional case, and so is not invariant under $U(k)$
subgroup of the rotational group.

In summary we have explored a class of matrix model which describes
identical particles in even dimensions. The number of free parameters of the
theory turns out to be the sum of that for individual particles.  The
dynamics is described by the moduli space metric on the blow-up space
of the  singular space $(C^k)^N/S_N$. We studied the
two particle moduli space in detail. We show that the minimum volume  of
$N$ particles grows like $N^k$.

Our model in two spatial dimensions and in four dimension with larger
number of flavor $M\ge 2$ does not have the direct interpretation as
the solitons in the field theory or the string theory. It would be
interesting to find such field theoretic model.  Finally, it would be
a challenge to find similar matrix model for supersymmetric D0 branes
on D8 branes with nonzero B field.

\subsection*{Acknowledgement}
This work is supported in part by the BK21 program of Ministry of
Education (C.K.,S.-H.Y.) and KOSEF 1998 interdisciplinary research 
grant 98-07-02-07-01-5 (K.L.). We appreciate  Dongsu Bak and Jeong-Hyuck Park
for many discussions on the subject. 

\newpage

\noindent{\bf Appendix A.}
\setcounter{equation}{0}
\renewcommand{\theequation}{A.\arabic{equation}} 

To find out the number of free parameters in Eqs.~(\ref{kq}) and
(\ref{fterm}), let us start by noting all commuting square matrices
can be triangularized simultaneously by unitary matrices. We
triangularize all $Z_i$ matrices.  After triangularization, we count
the number of free real parameters.To do that, we want to show that
$[Z_i,Z_{i+1}]=0 $ for $i=1,..,k-1$, then $[Z_i,Z_j]=0$ for arbitrary
$i,j$.  Let us first start by showing  that this is true for three
upper triangular complex matrices $Z_1,Z_2,Z_3$.

\noindent Lemma.  $[Z_1,Z_2]=[Z_2,Z_3]=0$ implies $[Z_3,Z_1]=0$ for
generic upper triangular matrices.

\noindent Proof) We prove this by induction.
Suppose the result holds for $ k \times  k$ matrices. Then for
$(k+1)\times (k+1)$
$Z_i$'s, we write 
\beq 
    Z_i = \left( \begin{array}{cc}
                     S_i  & v_i  \\
                    0  &  z_i  \\
                   \end{array} \right),
\eeq
where $S_i$ is a $k\times k$ upper triangular matrix, $v_i$ is a $k$
dimensional column vector and $z_i$ is a number. Then
\beq
[Z_i, Z_j] = \left( \begin{array}{cc} 
           [S_i, S_j]  &   S_i v_j + z_j v_i - S_j v_i - z_i v_j \\
                   0  &                    0                        \\
                 \end{array} \right) .
\eeq
Now from $[Z_1, Z_2] = [Z_2,Z_3]$ = 0 we have $[S_1, S_2] = [S_2, S_3] = 0.$
Then by induction hypothesis, it implies $[S_3,S_1] = 0.$ 
So the only thing need to prove is
\beq
         S_3 v_1 + z_1 v_3 - S_1 v_3 - z_3 v_1 = 0 .
\eeq
Indeed, this can be easily shown using the corresponding equations
from (1,2) and (2,3) pairs. More precisely, after short calculations
we get
\beq
  (S_2 - z_2) ( S_3 v_1 + z_1 v_3 - S_1 v_3 - z_3 v_1 ) = 0.
\eeq
But in general the determinant of $S_2 - z_2$ is not zero since
 $ \det(S_2 - z_2) = \prod_{a=1}^{k} (( S_{2})_{aa} - z_2)$
and the diagonal components of $S_2$ are not constrained by the commutators. 
Q.E.D.

For four triangular matrices such that
$[Z_1,Z_2]=[Z_2,Z_3]=[Z_3,Z_4]=0$, the above lemma shows that
$[Z_1,Z_3]=0$. Thus by using the above lemma again for $Z_1,Z_3,Z_4$,
we see $[Z_1,Z_4]=0$. By induction, then it should be true for
arbitrary number of $Z_i$'s.

For each triangular $Z_i$, there are $N(N+1)$ real parameters. The
total number of parameters of triangular $Z_i's$ and $\psi$ are
$kN(N+1) + 2NM$. The remaining gauge parameter after triangularization
is $N$ and Eq.(\ref{kq}) imposes $N^2$ constraints. Due to the above
argument there are $(k-1)N(N-1)$ constraints on $[Z_i,Z_{i+1}]=0$.  Then
total free parameters are
\beq
kN(N+1)+2NM - N^2-N-(k-1)N(N-1) =  N(2k + 2M-2) .
\eeq

\newpage

\noindent{\bf Appendix B.}
\setcounter{equation}{0}
\renewcommand{\theequation}{B.\arabic{equation}}

Here we explore the two dimensional case in detail. For $k=1,M=1$, the
general solutions of  the constraint (\ref{kq}) for any $N$ have been
found by introducing two Hermitian matrices $X, Y$ such that $ Z=
 (X+iY)/\sqrt{2} $.  After diagonalizing one of them by the
unitary transformation, we use the remnant $U(1)^N$ gauge freedom to
put the column vector $\psi$ to be real positive. The solution of the
constraint (\ref{kq}) is given as
\beqn
&& X_{ij}=\left \{ \begin{array} {cc} x_i & {\rm for}\;\; i=j  \\ 
           -i\zeta/(y_i-y_j)
             & {\rm for}\; \; i\neq j
       \end{array} \right. \; , \nonumber \\
&& Y_{ij}= y_i\delta_{ij}, \;\; \psi_i=\sqrt{\zeta}. 
\eeqn
In this coordinate the rotational symmetry is not manifest. (While the
diagonal components of the triangularized $Z$ matrices is rotationally
manifest and the distance between particles are obvious, it is much
hard to solve the constraint (\ref{kq}) in this coordinate.) Since the
general solution is obtained, we might say that the moduli space
metric can be obtained in principle and the multiparticle dynamics
could be understood.  However it turns out that it is technically
somewhat difficult.  Here we consider only the two particle case  and
find its explicit moduli space metric in this form of the
solution. Again the  Gauss law determines the infinitesimal gauge
parameter  $\delta \alpha$ uniquely. After some calculation, the
metric for the relative motion is given by
\beq
ds^2_{rel}=\frac{1}{8(x^2+y^2+{4\zeta^2\over y^2})}\left(
\biggl[x^2+y^2\biggr]dx^2+\bigg[x^2+{1\over y^2}(y^2+{4\zeta^2\over y^2})^2
\biggr]
dy^2-\frac{8\zeta^2x}{ y^3}dxdy 
\right), \\
\eeq
where $x\equiv x_1-x_2,\ \ y\equiv y_1-y_2$. (In this case it turns
out $\delta \psi = 0$, too.) 

Although $(x_i,y_i)$ can be identified as the position of particles at
large separation, its meaning at short distance changes.  We can
relate $(x,y)$ to the relative coordinate $z$ by taking the
eigenvalues of $Z=(X+iY)/\sqrt{2}$.  One can see that these two
coordinates $(x,y)$ for moduli space metric do not have one to one
correspondence though one form of the metric can be obtained from the
other by the direct coordinate relations.

We need to elucidate the Polychronakos system at this point because
Polychronakos has studied the first order time derivative
Lagrangian~\cite{poly}
\beq
 {\cal L}_1 = \kappa  \tr\left(  i\bar{Z} D_0 Z + i\bar{\psi} D_0
\psi-\zeta A_0\  \right) - w^2\tr \bar{Z} Z, 
\label{lag1} 
\eeq
where $D_0Z=\dot{Z}-i[A_0,Z] $, and $\ D_0\psi =\dot{\psi}-iA_0\psi.$.
The auxiliary field $A_0$ leads to the Gauss law constraint $
[Z,\bar{Z}] + \psi\bar{\psi} = \zeta 1_N $, which is identical to the
constraint (\ref{kq}) Classically there are $2N^2+2N $ degrees of
freedom in $Z$ and $\psi$.  Since there is only a first order term in
time derivative, the space of $Z,\psi$ forms a phase space.  $N^2$
parameters are constrained by the above Gauss law and another $N^2$
parameters can be absorbed to the gauge parameters, leaving free $2N $
parameters in the phase space.  These $2N$ parameters denote the $N$
particle location on two dimensional noncommutative plane. The
eigenvalues and eigenstates of the Hamiltonian $H= w^2 \tr(\bar{Z}Z )$
are completely solved~\cite{heller}.

The first order kinetic energy (\ref{lag1}) leads to one form $A=\tr( Z
d\bar{Z} + \psi d \bar{\psi})$ on the $2N$ dimensional
moduli space, making it to be a phase space. As $dA= \tr(\delta
Z\wedge \delta \bar{Z} + \delta \psi\wedge \delta \bar{\psi}) $
regardless of $\delta \alpha$, the K\"ahler two form of our moduli space
gives the symplectic structure of the Polychronakos system.  Our
moduli space becomes the phase space of the first order system. It
would be interesting to work out the quantum problem in terms of the
moduli space coordinate.

\end{document}